\documentclass{article}
\usepackage{spconf,amsmath,graphicx,hyperref}
\usepackage{tabularx} 
\usepackage{multirow}
\usepackage{amsfonts}
\usepackage{booktabs}
\usepackage{multirow}
\usepackage{siunitx}
\newcolumntype{Y}{>{\raggedright\arraybackslash}X} 
\usepackage[table]{xcolor}
\usepackage{amssymb} 
\usepackage{arydshln}
\usepackage{bm} 
\usepackage{setspace}   
\usepackage[numbers,sort&compress]{natbib}
\usepackage{adjustbox}

\newcommand{\newpara}[1]{\vspace{1mm}\noindent\textbf{#1}}



\usepackage[hang,flushmargin]{footmisc}

\usepackage[font=small,skip=2pt]{caption}
\title{Diffusion-Link: Diffusion Probabilistic Model \\for Bridging the Audio-Text Modality Gap}
\name{KiHyun Nam$^{1*}$, Jongmin Choi$^{1*}$, Hyeongkeun Lee$^1$, Jungwoo Heo$^2$,  Joon Son Chung$^1$\thanks{$^*$These authors contributed equally to this work.}}
\address{$^1$Korea Advanced Institute of Science and Technology, South Korea, $^2$University of Seoul, South Korea}
\begin{document}
\ninept
\maketitle
\begin{abstract}
Contrastive audio–language pretraining yields powerful joint representations, yet a persistent audio–text modality gap limits the benefits of coupling multimodal encoders with large language models (LLMs). We present Diffusion-Link, a diffusion-based modality-bridging module that generatively maps audio embeddings into the text-embedding distribution. The module is trained at the output embedding from the frozen multimodal encoder and implemented as a lightweight network with three residual MLP blocks. To assess the effect of Diffusion-Link on multimodal encoder-LLM coupling, we evaluate on Automatic Audio Captioning (AAC); to our knowledge, this is the first application of diffusion-based modality bridging to AAC. We report two results. (1) Modality-gap analysis: on similarity and geometric criteria, Diffusion-Link reduces the modality gap the most among prior diffusion-based methods and shows a collective migration of audio embeddings toward the text distribution. (2) Downstream AAC: attaching Diffusion-Link to the same multimodal LLM baseline achieves state-of-the-art on AudioCaps in both zero-shot and fully supervised captioning without external knowledge, with relative gains up to 52.5$\%$ and 7.5$\%$, respectively. These findings show that closing the modality gap is pivotal for effective coupling between multimodal encoders and LLMs, and diffusion-based modality bridging offers a promising direction beyond knowledge-retrieval-centric designs. Code will be released upon acceptance\footnote{Official code: \href{https://github.com/DevKiHyun/Diffusion-Link}{{https://github.com/DevKiHyun/Diffusion-Link}}}. 
\end{abstract}

\vspace{-2mm}
\begin{figure*}[t!]
    \centering
    \includegraphics[width=0.95\linewidth]{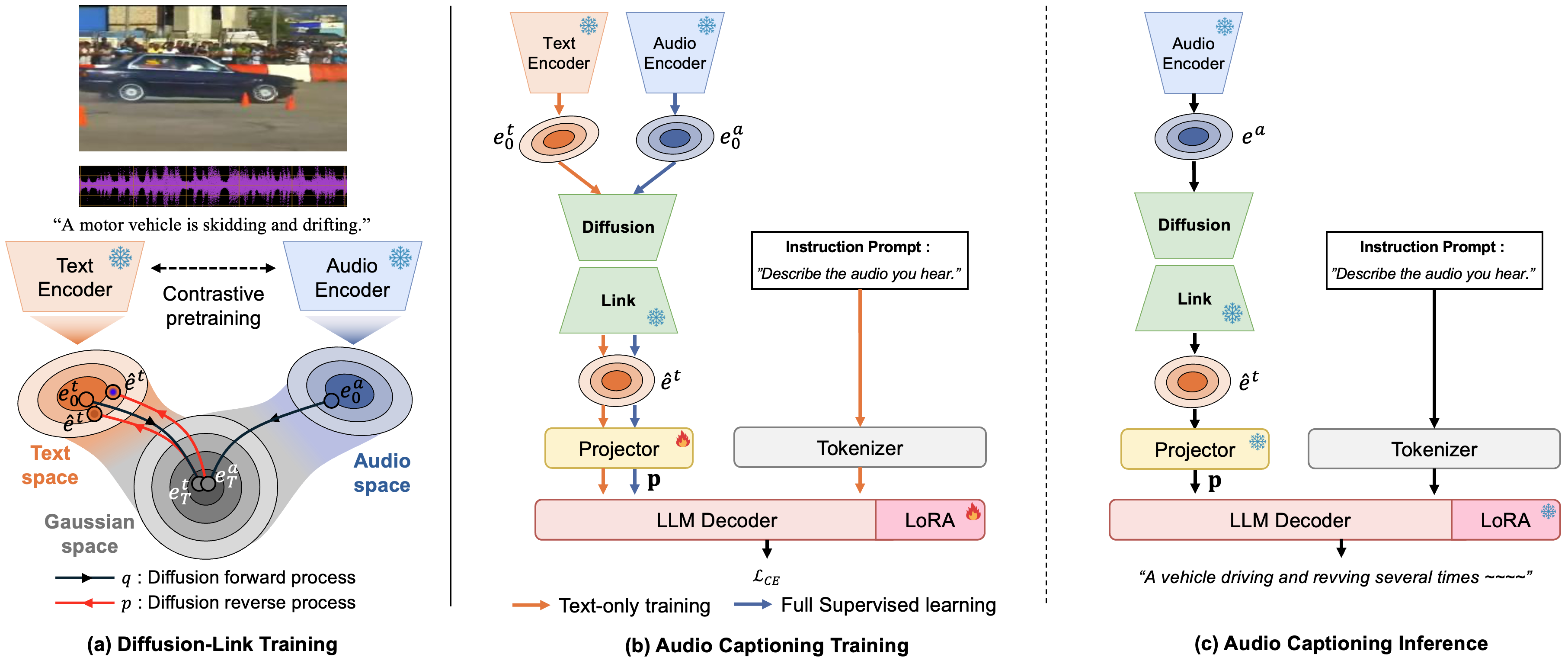}
    \caption{(a) Overview of the proposed Diffusion-Link mechanism and (b,c) illustration of our LLM-based AAC system with Diffusion-Link.}
    \label{fig:main_figures}
\end{figure*}

\section{Introduction}

Large-scale audio--language models have shown strong multimodal performance across a range of multimodal tasks.
In particular, CLAP~\cite{elizalde2023clap,wu2023laionclap} maps natural-language descriptions and acoustic signals into a shared embedding space via contrastive learning, achieving state-of-the-art results on various audio--language multimodal tasks~\cite{ghosh2025reclap}. In parallel, advances in LLMs~\cite{chiang2023vicuna, touvron2023llama, chung2024scaling} enable coupling contrastive audio--language encoders with powerful decoders, already demonstrating compelling audio–language reasoning and captioning~\cite{tang2023salmonn, rho2025lavcap}.

Yet recent studies reveal a structural modality gap in contrastive multimodal encoders. Liang et al.~\cite{NEURIPS2022_702f4db7} quantified the gap and linked its magnitude to zero-shot performance and fairness, while Zhang et al.~\cite{ICLR2024_e26f31de} analyzed embedding geometry and showed that gap reduction benefits cross-modal tasks. From an application angle, linking contrastive spaces~\cite{radford2021learning, elizalde2023clap} via mediating modalities enables unpaired transfer~\cite{wang2023connecting}, and broader alignment across audio–vision–text–3D yields competitive zero-shot results~\cite{NEURIPS2024_a71df365}. Taken together, these prior works suggest that addressing the modality gap is essential for improving zero-shot and cross-modal task performance.

Diffusion models~\cite{NEURIPS2020_4c5bcfec, song2021denoising} have become a standard generative paradigm in various fields, reliably producing high-fidelity samples~\cite{ramesh2022hierarchical, rombach2022high, liu2023audioLDM}. They learn a forward noising process toward an isotropic Gaussian and a reverse denoising process back to the target distribution. Viewing embedding vector as data, diffusion can learn a trajectory that bridges the embedding distributions between two modalities. 
We adopt this view and design a reverse process that first moves audio embeddings to a shared isotropic Gaussian waypoint and then maps them into the text-embedding distribution, thereby enabling effective modality bridging.

Recent embedding-generative works support this view. In speaker recognition, SEED~\cite{nam25b_interspeech} applies the forward process to both clean and noisy speaker embeddings and trains the reverse process to regenerate the clean speaker embeddings, introducing cross-sample prediction and demonstrating embedding-level generation. In vision--language, Diffusion-Bridge~\cite{Lee_2025_CVPR} trains only on CLIP text embeddings and injects image embeddings at an intermediate reverse step to convert them into text-like vectors--an early instance of embedding-space modality bridging.

We propose \textbf{Diffusion-Link}, which directly bridges the audio--text modality gap, building on prior works~\cite{nam25b_interspeech,Lee_2025_CVPR}. The key idea is to (i) use paired audio--text embeddings from an audio-language multimodal encoder during training to explicitly connect the two distributions, and (ii) achieve modality bridging by enforcing that the reverse process always map to the text embedding distribution. To this end, we gradually inject Gaussian noise into both embeddings in the forward process to send them to a common isotropic Gaussian state, and train with an L2 reconstruction loss so that the reverse process consistently predicts embeddings from the text distribution. Moreover, we add a topology loss that preserves the relative geometry of the text distribution by matching the within-batch cosine similarity structure of the original text and the generated text-like embeddings. At inference, Diffusion-Link outputs a text-like embedding regardless of the input modality. Diffusion-Link is a lightweight network composed of three residual multilayer perceptron (MLP) blocks, and the multimodal encoder is frozen during training. For practical validation, we attach Diffusion-Link after multimodal encoder as a plug-in and combine it with a LLM-based decoder to evaluate audio captioning. To our knowledge, this is the first attempt to apply diffusion-based modality bridging to audio captioning.

We verify consistent gains on the AudioCaps~\cite{audiocaps} dataset along two axes: modality-gap analysis and LLM-based downstream tasks. On similarity and geometric criteria, Diffusion-Link increases the similarity of paired audio--text samples  while decreasing that of unpaired, \textbf{achieving the largest gap reduction} over prior methods. Visualizations further show a clear collective migration of audio embeddings toward the text-embedding distribution after the diffusion process. In Automatic Audio Captioning (AAC), attaching Diffusion-Link as a plug-in to the same multimodal LLM baseline yields relative improvements of up to \textbf{52.5\%} in zero-shot audio captioning and \textbf{7.5\%} in fully supervised audio captioning, reaching \textbf{state-of-the-art in both cases} without external knowledge. Because many existing systems, especially in zero-shot, rely on external knowledge such as retrieval-augmented generation (RAG), these results establish Diffusion-Link as a new powerful solution that achieves consistent gains on the same multimodal LLM system while shifting the source of performance from knowledge retrieval to modality bridging.

\vspace{-4mm}
\section{Method}
\vspace{-1mm}
In this section, we describe the proposed framework (Fig.~\ref{fig:main_figures}). We denote by $\mathbf{e}^{a}_0, \mathbf{e}^{t}_0 \in \mathbb{R}^d$ the paired audio and text embeddings obtained from a multimodal encoder~\cite{elizalde2023clap}. For brevity, we use $\mathcal{M}$ to indicate the modality, with $\mathcal{M}$ representing audio $a$ and text $t$.

\vspace{-2mm}
\subsection{Background on Diffusion Probabilistic Models}
\label{sec:background-dpm}

We briefly review denoising diffusion probabilistic models (DDPM)~\cite{NEURIPS2020_4c5bcfec} under sample-prediction formulation. 

The forward diffusion process progressively corrupts a given sample $\mathbf{z}_0 \sim q(\mathbf{z}_0)$ at each timestep $s=1,\dots,T$:
\begin{equation}
    q(\mathbf{z}_s|\mathbf{z}_0)
    = \mathcal{N}\!\big(\mathbf{z}_s;\sqrt{\bar{\alpha}_s}\,\mathbf{z}_0,(1-\bar{\alpha}_s)\mathbf{I}\big),
    \label{eq:forward}
\end{equation}
where $0\leq\alpha_s\leq 1$ is the noise schedule, $\bar{\alpha}_s=\prod_{\tau=1}^s \alpha_\tau$, and $\mathbf{I}$ is the identity matrix. This also admits the following closed-form reparameterization:
\begin{equation}
    \mathbf{z}_s=\sqrt{\bar{\alpha}_s}\mathbf{z}_0+\sqrt{1-\bar{\alpha}_s}\,\boldsymbol{\epsilon},
    \quad \boldsymbol{\epsilon}\!\sim\!\mathcal{N}(\mathbf{0},\mathbf{I}).
\end{equation}

The reverse diffusion process gradually denoises $\mathbf{z}_t$ back toward the data distribution at each timestep $s$:
\begin{equation}
    p_\theta(\mathbf{z}_{s-1}|\mathbf{z}_s)
    = \mathcal{N}\!\big(\mathbf{z}_{s-1};\,\mu_\theta(\mathbf{z}_s,s), \sigma_s^2 \mathbf{I}\big),
    \label{eq:reverse}
\end{equation}
where $\mu_\theta(\mathbf{z}_s,s)$ is parameterized by a neural denoiser. The denoiser $\phi_\theta(\cdot,s)$ is trained to predict the sample $\mathbf{z}_0$ at $s=0$ via the objective
\begin{equation}
    \mathcal{L}
    = \mathbb{E}_{\mathbf{z}_0,s,\boldsymbol{\epsilon}}
    \big\lVert \mathbf{z}_0 - \phi_\theta(\mathbf{z}_s,s) \big\rVert_2^2 .
    \label{eq:x0-loss}
\end{equation}

\vspace{-2mm}
\subsection{Modality Gap Bridging via Diffusion-Link}
\vspace{-1mm}
\label{sec:method-dlink}

Diffusion-Link is a neural network denoiser trained at the output embeddings of the multimodal encoder.

\subsubsection{Training Objective}
\label{sec:training-dlink}

We apply the same forward process \eqref{eq:forward} to each modality $\mathcal{M}$:
{\small
\begin{equation}
\mathbf{e}^\mathcal{M}_s=\sqrt{\bar{\alpha}_s}\,\mathbf{e}^\mathcal{M}_0
+\sqrt{1-\bar{\alpha}_s}\,\boldsymbol{\epsilon}_\mathcal{M},
\quad
\boldsymbol{\epsilon}_\mathcal{M}\sim\mathcal{N}(\mathbf{0},\mathbf{I}).
\label{eq:forward-bimodal}
\end{equation}
}
The denoiser $\phi_\theta(\cdot,s)$ is trained under the sample-prediction formulation to map both noised text and audio embeddings to the \emph{text} embedding distribution at $s\!=\!0$.  
This yields the cross-sample prediction loss~\cite{nam25b_interspeech}:
{\small
\begin{equation}
\mathcal{L}_{\text{diff}}
= \mathbb{E}\Big[
    \underbrace{\lVert \mathbf{e}^t_0 - \phi_\theta(\mathbf{e}^t_s,s) \rVert_2^2}_{\text{text}\to\text{text}}
    + \underbrace{\lVert \mathbf{e}^t_0 - \phi_\theta(\mathbf{e}^a_s,s) \rVert_2^2}_{\text{audio}\to\text{text}}
\Big],
\label{eq:dlink-loss}
\end{equation}
}
where the first term enforces high-fidelity reconstruction of text-like embeddings, while the second term encourages audio embeddings toward the text distribution.

Furthermore, we introduce a batch-level topology loss to preserve the relative geometry of the text distribution.
Let $\mathbf{X}\!=\![\mathbf{e}^t_{0,i}]_{i\in\mathcal{B}}$ and 
$\hat{\mathbf{X}}\!=\![\hat{\mathbf{e}}^t_{i}]_{i\in\mathcal{B}}$ 
denote the text and text-like embedding matrices.
Row-wise $\ell_2$-normalized matrices $\mathbf{X}'$ and $\hat{\mathbf{X}}'$ are obtained from $\mathbf{X}$ and $\hat{\mathbf{X}}$, respectively, yielding similarity matrices 
$\mathbf{S}_{xx}=\mathbf{X}'\mathbf{X}'^\top \in \mathbb{R}^{\mathcal{B}\times\mathcal{B}}$ and 
$\mathbf{S}_{x\hat x}=\mathbf{X}'\hat{\mathbf{X}}'^\top\in \mathbb{R}^{\mathcal{B}\times\mathcal{B}}$, 
and the topology loss is the squared Frobenius distance:
\begin{equation}
\mathcal{L}_{\text{topo}}
=\|\mathbf{S}_{xx}-\mathbf{S}_{x\hat x}\|_{F}^{2}.
\label{eq:topo}
\end{equation}

The total training objective is  
\begin{equation}
\mathcal{L}_{\text{total}}=\mathcal{L}_{\text{diff}}+\mathcal{L}_{\text{topo}}.
\end{equation}
\subsubsection{Inference to generate text-like embedding}
\label{sec:inference-dlink}
At inference, given $\mathbf{e}^\mathcal{M}$, we optionally apply forward noising at step $s_\ast$ with $\boldsymbol{\epsilon}\!\sim\!\mathcal{N}(\mathbf{0},\mathbf{I})$, and then run the learned reverse trajectory to $s\!=\!0$ using DDIM sampler~\cite{song2021denoising}:
\begin{align}
\mathbf{e}^\mathcal{M}_{s_\ast}\!&=\!\sqrt{\bar{\alpha}_{s_\ast}}\,\mathbf{e}^\mathcal{M}\!+\!\sqrt{1-\bar{\alpha}_{s_\ast}}\,\boldsymbol{\epsilon}, \quad \textit{(Forward)}\\
\hat{\mathbf{e}}^t\!&=\!\text{DDIM}\big(\phi_\theta,\mathbf{e}^\mathcal{M}_{s_\ast},\,s_\ast \!\to\! 0\big), \quad \textit{(Reverse)}
\end{align}
The output $\hat{\mathbf{e}}^t$ is a \emph{text-like} embedding.

\begin{figure}[t!]
    \centering
    \includegraphics[width=0.8\linewidth]{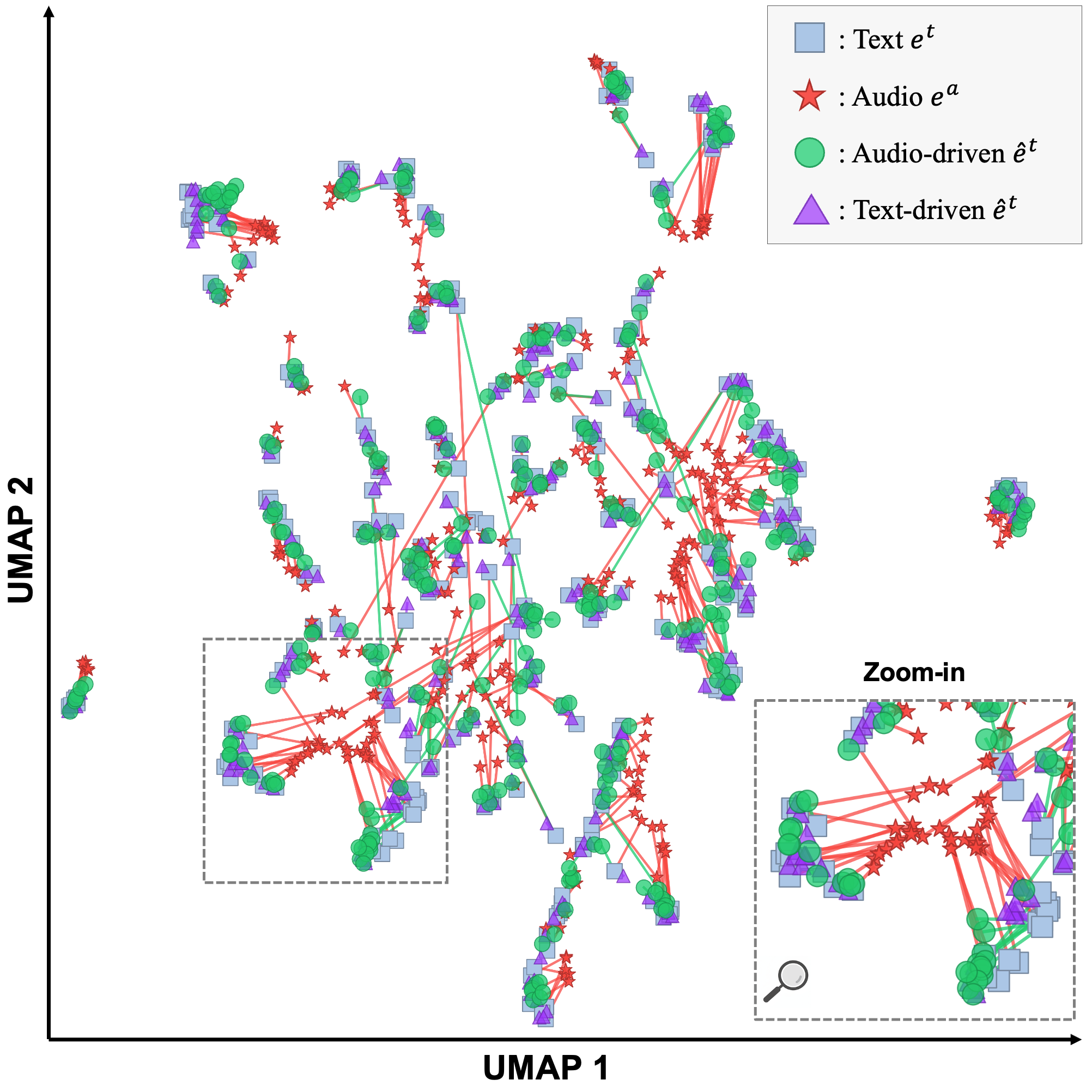}
    \caption{Visualization of embeddings on AudioCaps using UMAP. Red line means the pair of audio and text embeddings. Green line means the pair of text-like and original text embeddings.}
    \vspace{-6mm}
    \label{fig:visualization}
\end{figure}

\subsection{LLM-based Text Decoding}
\label{sec:decoder}

Given a text-like embedding $\hat{\mathbf{e}}^t$, a projection head maps it to a soft-prefix vector $\mathbf{p}\in\mathbb{R}^{mh}$. Here $m$ denotes the number of learnable soft tokens and $h$ denotes the decoder hidden size. We then $\mathbf{p}$ into soft-prefix tokens sequence $\mathbf{p}\in\mathbb{R}^{m\times h}$ and feed this sequence to the decoder. If we optionally prepend a fixed instruction prompt of $n$ tokens, the resulting input becomes $\mathbf{p}\in\mathbb{R}^{(m+n)\times h}$.

We consider two training options for our LLM decoder framework:
\textbf{(i) text-only training}: using text-driven $\hat{\mathbf{e}}^t$ with an instruction prompt.
\textbf{(ii) fully supervised training}: using audio-driven $\hat{\mathbf{e}}^t$ and we not use an instruction prompt.
We train the LLM decoder with the standard autoregressive cross-entropy objective
\begin{equation}
\mathcal{L}_{\text{CE}}
= - \sum_{l=1}^{L} \log p_\psi\big(w_l|w_{<l}, \mathbf{p}\big),
\label{eq:ar-loss}
\end{equation}
where $\psi$ denotes the LLM decoder's learnable parameters and $\mathbf{w}=(w_1,\dots,w_L)$ means the target caption tokens.
At inference, given audio data only with optional instruction prompt, and decode target caption.
When the LLM decoder is trained under the text-only training, this evaluation corresponds to zero-shot captioning.

\vspace{-2mm}
\section{Experimental Settings}
\label{sec:exp-settings}

\subsection{Datasets}
\label{sec:datasets}
For training and evaluation, we conduct all experiments on AudioCaps~\cite{audiocaps}, a corpus of ten-second audio clips paired with human-written captions. We use 48,595 training clips and 944 test clips. The \textit{train} split provides one caption per clip, whereas the \textit{test} split provides up to five. All audio is resampled to 48kHz. For audio preprocessing, we compute STFTs with a 1,024 window size and a 480 hop length, and then form mel-spectrograms with 64 mel-bins. We train on the \textit{train} split and report results on the \textit{test} split.

\sisetup{
  table-format = 1.3,
  table-number-alignment = center,
  detect-weight = true,
  detect-inline-weight = math
}
\newcolumntype{C}{>{\centering\arraybackslash}X}

\begin{table}[t!]
\centering
\caption{Average cosine similarity scores for various embedding pairs on AudioCaps. For the CLAP, no transformation is applied, so $\hat e^\mathcal{M}\!=\!e^\mathcal{M}$. We report $e^t\!\cdot\!\hat e^{t\leftarrow \mathcal{M}}$, where $\hat e^{t\leftarrow \mathcal{M}}$ denotes a text-like embedding generated from modality $\mathcal{M}$. Transformations are obtained via C3~\cite{ICLR2024_e26f31de}, DB (Diffusion-Bridge)~\cite{Lee_2025_CVPR}, DG (DiffGap)~\cite{mo2025diffgap}, and our DL (Diffusion-Link). Here, $\sim$ and $\not\sim$ indicate matched and non-matched pairs, respectively.}
\label{tab:diffusion_cosine}
\begingroup
\footnotesize
\setlength{\tabcolsep}{5pt}   
\renewcommand{\arraystretch}{0.95}

\begin{tabularx}{0.9\linewidth}{@{}C *{5}{S}@{}} 
\toprule
\multicolumn{1}{c}{Comparison Pair} & \multicolumn{5}{c}{Cosine Similarity} \\
\cmidrule(lr){2-6}
& {CLAP} & {C3} & {DB} & {DG} & {DL} \\
\midrule
(a) $e^{t}\!\cdot\!\hat{e}^{t\leftarrow a}(\sim)$ & 0.486 & 0.547 & 0.528 & 0.110 & {\bfseries 0.688} \\
(b) $e^{t}\!\cdot\!\hat{e}^{t\leftarrow t}(\sim)$  & {\bfseries 1.000} & 1.000 & 0.999 & 0.334 & 0.945 \\
(c) $e^{t}\!\cdot\!\hat{e}^{t\leftarrow a}(\not\sim)$ & 0.030 & 0.092 & {\bfseries 0.000} & 0.007 & {\bfseries 0.000} \\
(d) $e^{t}\!\cdot\!\hat{e}^{t\leftarrow t}(\not\sim)$ & 0.098 & 0.158 & 0.002 & 0.043 & {\bfseries 0.001} \\
\bottomrule
\end{tabularx}
\endgroup
\end{table}
\begin{table}[t!]
\centering
\vspace{-4mm}
\caption{Average cosine similarity scores for various inference forward timestep $s_\ast$ during diffusion process on AudioCaps.}
\label{tab:diffusion_sampling}
\resizebox{0.9\linewidth}{!}{
\begin{tabular}{lccccc}
\toprule
\multicolumn{1}{c}{\multirow{2.4}{*}{Diffusion-Link}} & \multicolumn{5}{c}{Inference forward timestep $s_\ast$} \\
\cmidrule(lr){2-6}
& 100 & 200 & 300 & 400 & 500 \\
\midrule
Cosine Similarity &  \textbf{0.688}  &   0.654   &   0.596   &  0.510   &  0.404  \\
\bottomrule
\vspace{-3mm}
\end{tabular}
}
\end{table}
\sisetup{table-format=2.1, detect-weight=true, detect-inline-weight=math}
\newcolumntype{Y}{>{\raggedright\arraybackslash}X} 
\newcolumntype{C}{>{\hsize=0.6\hsize\centering\arraybackslash}X}

\begin{table*}[t]
\centering
\caption{Performance comparison of AAC models on AudioCaps. \textbf{External knowledge \#} is the number of non-audio samples used by the LLM at test time. For a fair comparison on the embedding-level modality-gap problem, $^\dagger$ results use only embedding-level RAG without external $k$-caption selection.}
\label{tab:performance_captioning}

\begingroup
\footnotesize
\setlength{\tabcolsep}{4pt}
\renewcommand{\arraystretch}{0.95}

\begin{tabularx}{0.7\linewidth}{@{\extracolsep{\fill}}X C C S S S S@{}}
\toprule
\textbf{Method} &
\textbf{\begin{tabular}[c]{@{}c@{}}Encoder \\output dim.\end{tabular}} &
\textbf{\begin{tabular}[c]{@{}c@{}}External \\knowledge \#\end{tabular}} &
\multicolumn{1}{c}{\textbf{ME}\(\uparrow\)} &
\multicolumn{1}{c}{\textbf{CD}\(\uparrow\)} &
\multicolumn{1}{c}{\textbf{SP}\(\uparrow\)} &
\multicolumn{1}{c}{\textbf{SD}\(\uparrow\)} \\
\midrule
\textit{\textbf{Zero-shot Captioning}} & & & & & & \\
ZerAuCap~\cite{salewski2023zero}            & $1\times D$ & 527       & 12.3 & 28.1 &  8.6 & 18.3 \\
DRCap$^\dagger$~\cite{li2025drcap}          & $1\times D$ & 450{,}000 & 21.8 & 59.5 & 15.7 & 37.6 \\
Zhang \textit{et al.}~\cite{zhang2025zero}  & $1\times D$ & No        & 22.0 & 64.4 & 15.6 & 40.0 \\
WSAC~\cite{Kouzelis2023}                    & $1\times D$ & 46{,}000  & 24.1 & 63.3 & 17.3 & 40.3 \\
\hdashline
\textbf{Ours}                               & $1\times D$ & No        & \textbf{24.2} & \textbf{73.2} & \textbf{17.5} & \textbf{45.4} \\
\midrule
\textit{\textbf{Fully Supervised Captioning}} & & & & & & \\
Prefix AAC~\cite{kim2023prefix}             & $T\times D$ & No        & 24.0 & 73.3 & 17.7 & 45.5 \\
RECAP~\cite{ghosh2024recap}                 & $T\times D$ & 600{,}000 & 25.6 & 75.1 & 18.6 & 47.1 \\
EnCLAP-large~\cite{kim2024enclap}           & $T\times D$ & No        & 25.5 & 80.3 & 18.8 & 49.5 \\
CLAP-ART~\cite{takeuchi2025clap}            & $T\times D$ & No        & 25.6 & 80.7 & 18.8 & 49.8 \\
\hdashline
\textbf{Ours}                               & $1\times D$ & No        & \textbf{25.6} & \textbf{82.5} & \textbf{18.9} & \textbf{50.7} \\
\bottomrule
\end{tabularx}
\endgroup
\end{table*}

\begin{table}[t]
\centering
\caption{Ablation study to analyze the effectiveness of diffusion-based modality bridging method.}
\label{tab:ablation_captioning}

\begingroup
\small                         
\setlength{\tabcolsep}{3pt}   
\renewcommand{\arraystretch}{0.88}

\begin{tabularx}{0.9\linewidth}{@{}Y S[table-format=2.1] S[table-format=2.1] S[table-format=2.1] S[table-format=2.1]@{}}
\toprule
\textbf{Method} & 
\textbf{ME}$\uparrow$ & \textbf{CD}$\uparrow$ & \textbf{SP}$\uparrow$ & \textbf{SD}$\uparrow$ \\
\midrule
\textit{\textbf{Zero-shot Captioning}} \\
Baseline (CLAP \& LLaMa2-7B)  & 21.2 & 48.0 & 14.4 & 31.2 \\
\quad + Diffusion-Bridge~\cite{Lee_2025_CVPR} & 23.3 & 62.6 & 16.5 & 39.5 \\
\quad + \textbf{Diffusion-Link (Ours)} & \textbf{24.2} & \textbf{73.2} & \textbf{17.5} & \textbf{45.4} \\
\midrule
\multicolumn{5}{@{}l@{}}{\scalebox{0.99}{\textit{\textbf{Fully Supervised Captioning}}}}\\[-1pt]
Baseline (CLAP \& LLaMa2-7B) & 25.0 & 76.9 & 18.6 & 47.7 \\
\quad+ Diffusion-Bridge~\cite{Lee_2025_CVPR} & 25.2 & 77.1 & 18.0 & 47.4 \\
\quad + \textbf{Diffusion-Link (Ours)} & \textbf{25.6} & \textbf{82.5} &  \textbf{18.9} & \textbf{50.7} \\
\bottomrule
\end{tabularx}
\endgroup
\end{table}

\subsection{Implementation Details and Metrics}

For audio-language multimodal encoder, we use the LAION-CLAP pretrained model~\cite{wu2023laionclap} and keep it frozen.
Following prior work~\cite{Lee_2025_CVPR}, we apply the same normalization process to the output embeddings of CLAP. For Diffusion-Link, we adopt three residual MLP blocks~\cite{li2024return}. We train Diffusion-Link with the Adam~\cite{kingma20153rd} optimizer and a batch size of $128$. The base learning rate is set to $1{\times}10^{-4}$ and follows a step-decay schedule, multiplying the rate by $0.97$ every $200$ steps. We employ an exponential moving average (EMA) of the model parameters with a decay of $0.995$ and use the EMA weights for inference. We adopt a cosine noise schedule with a total of $T{=}1000$ timesteps.
At inference, we employ DDIM~\cite{song2021denoising} sampling with 5 iteration steps. Before denoising, we apply a shallow forward noising to $s_\ast{=}100$ and then run the reverse procoess. For LLM-based text decoder, we adopt LLaMA2(7B)~\cite{touvron2023llama} as the LLM decoder. In the text-only training, we employ a linear layer with soft prefix tokens $m{=}1$ for the project head and prepend a short instruction prompt; in the fully supervised training, we use 2 linear layers with $m{=}10$ for the project head  and no hard prompt. We fine-tune project head and the LLM using LoRA~\cite{hu2022lora}. LLM training uses AdamW~\cite{loshchilov2018decoupled} optimizer with batch size $4$ for $50$ epochs: the learning rate warms up over the first $2$ epochs with max learning rate $5{\times}10^{-6}$, then use a cosine decay. We also train a baseline multimodal LLM system to verify the effectiveness of Diffusion-Link, we adopt same setting but detach only Diffusion-Link module. For evaluation, we adopt the metrics for modality gap analyzing, including cosine simiarity and visualization using UMAP~\cite{McInnes2018}. For AAC, we use the metrics, METEOR (ME)~\cite{banerjee2005meteor}, CIDEr (CD)~\cite{vedantam2015cider}, SPICE (SP)~\cite{anderson2016spice}, and SPIDEr (SD)~\cite{liu2017improved}.

\vspace{-2mm}
\section{Results}
\label{sec:results}
\vspace{-2mm}

\subsection{Main Results}
\newpara{Effectiveness of Diffusion-Link for Modality Bridging.}
As shown in Table~\ref{tab:diffusion_cosine}, Diffusion-Link attains the highest cosine similarity on matched audio--text pairs.
While most approaches improve over CLAP, DiffGap underperforms, because it generates from pure Gaussian noise with the input embedding condition, which weakens information reconstruction.
By contrast, Diffusion-Link treat the input embedding as residing at an intermediate reverse step, thereby minimizing information loss and ensuring high-quality generation along the reverse trajectory.
Importantly, Diffusion-Link also yields the lowest similarity on non-matched pairs, indicating not merely a global contraction of the space but maintaining semantic information. 
Figure ~\ref{fig:visualization} visualizes this effect. Both the generated text-like embeddings from audio and text embeddings all move toward the ground-true text embedding distribution, demonstrating that Diffusion-Link has learned a stable generative modality bridge for the text embedding distribution, regardless of the input modality.

\newpara{Diffusion-Link Amplifies Multimodal Encoder-LLM Coupling.}

Table~\ref{tab:performance_captioning} compares a range of AAC systems. In contrast to many prior methods that leverage longer audio representations or external knowledge (e.g., RAG), our multimodal LLM system captures input audio feature with only a single \(1{\times}D\) text-like embedding produced by Diffusion-Link, and achieves SOTA in both zero-shot and fully supervised captioning \emph{without external knowledge}. Notably, considering that most prior zero-shot models rely heavily on external knowledge, outperforming them without any external knowledge demonstrates the significant efficiency of Diffusion-Link.

According to Table~\ref{tab:ablation_captioning}, our baseline LLM-based AAC system is not competitive relative to prior AAC systems in Table~\ref{tab:performance_captioning}. However, applying Diffusion-Link markedly improves the same backbone. In zero-shot audio captioning, we observe a \textbf{52.5$\%$} relative increase in CIDEr together with substantial gains on the other metrics. These dramatic gains demonstrate that Diffusion-Link is the key factor and reaffirm the primacy of \emph{modality-gap reduction} over using longer audio representations or external knowledge. Moreover, in fully supervised audio captioning we observe up to \textbf{7.3$\%$} relative improvement, underscoring our method’s applicability.

\subsection{Ablation Studies}
We conduct ablations to analyze how the depth of forward noising affects modality bridging and high-quality generation. According to Table~\ref{tab:diffusion_sampling}, increasing the inference forward timestep \(s_\ast\) from shallow levels initially keeps similarity quite stable; beyond a threshold, the similarity drops sharply as \(s_\ast\) increases. This indicates that over-noising pushes representations deeper into the common Gaussian space and \emph{erases information}, thereby degrading semantic preservation in the reconstructed text-like embeddings. 

This finding is consistent in AAC results. In Table~\ref{tab:diffusion_cosine}, the similarity score of Diffusion-Bridge is similar to that of Diffusion-Link when \(s_\ast\) is between 300 and 400. This suggests that the performance of Diffusion-Bridge corresponds to over-noising of Diffusion-Link, which aligns with the observed \emph{semantic information loss}. Furthermore, under the same multimodal LLM system, the AAC results in Table~\ref{tab:ablation_captioning} follow the same pattern: attaching Diffusion-Link yields large gains, whereas using Diffusion-Bridge provides only limited improvements. Together, the three tables show that \emph{excessive} forward noising reduces similarity and weakens bridging, which in turn harms downstream performance; conversely, choosing an appropriate \(s_\ast\) maximizes content preservation in the text-like embedding, strengthens conditioning-distribution alignment for the LLM decoder, and translates into AAC gains.

\vspace{-2mm}
\section{Conclusion}
\vspace{-2mm}
We introduced \textbf{Diffusion-Link}, a lightweight residual MLP diffusion module that bridges audio embeddings to the text embedding distribution without keeping the multimodal encoder frozen. The method aligns the conditioning input by increasing matched similarity and decreasing mismatched similarity. On AAC, it improves the same multimodal LLM baseline by \textbf{52.5\%} and \textbf{7.3\%} without external knowledge for zero-shot and fully supervised AAC, respectively. This plug-in-play approach of Diffusion-Link is expected to generalize beyond audio captioning and enable effective zero-shot performance in a variety of multimodal LLMs.

\clearpage

\setlength{\bibsep}{1pt}
\setstretch{0.8}
\bibliographystyle{IEEEbib}
\bibliography{shortstrings,refs}

\end{document}